\documentclass[aps,prb,preprint,superscriptaddress,amssymb,amsmath]{revtex4-1}
\usepackage{graphicx}
\usepackage{dcolumn}
\usepackage{bm, color}
\usepackage[none]{hyphenat} 
\usepackage{amsmath}

\begin{document}

\title{(Quasi-)Quantization of the electrical, thermal, and thermoelectrical conductivities in two and three dimensions}
\author{Jonathan Noky}
\affiliation{Max Planck Institute for Chemical Physics of Solids, D-01187 Dresden, Germany}
\author{Johannes Gooth}
\affiliation{Max Planck Institute for Chemical Physics of Solids, D-01187 Dresden, Germany}
\author{Yan Sun} 
\affiliation{Max Planck Institute for Chemical Physics of Solids, D-01187 Dresden, Germany}
\author{Claudia Felser} 
\email{Claudia.Felser@cpfs.mpg.de}
\affiliation{Max Planck Institute for Chemical Physics of Solids, D-01187 Dresden, Germany}

\date{\today}

\begin{abstract}
The quantum Hall effect in a 2D electron system expresses a topological invariant, leading to a quantized conductivity. The thermal Hall and thermoelectric Nernst conductances in two dimensions are also reported to be quantized in specific systems. However, a comprehensive study of these quantities within one formalism for quantum Hall systems is so far elusive. In this work, we investigate the Hall, thermal Hall, and Nernst effects analytically and numerically in 2D and 3D. In addition to the quantized values for the Hall and thermal Hall conductances in two dimensions, we also find quasi-quantized values in three dimensions, which are the related 2D quanta scaled by a characteristic length. The Nernst conductivity is not generally quantized. Instead, an integration in energy is required to obtain a universally (quasi-)quantized thermoelectric quantity.
\end{abstract}
\maketitle

\section{Introduction}

Topological states of matter are characterized by topological invariants, which are expressed through physical quantities whose values are quantized and do not depend on the details of the system (such as its shape, size, and impurities). The best-known of these quantities is the Hall conductivity $\sigma$, which is the ratio of a transverse electrical current, generated by a longitudinal voltage under broken time-reversal symmetry (TRS)~\cite{hall1879new}. When electrons are confined to a two-dimensional (2D) system and are subjected to an external magnetic field $B$, $\sigma$ becomes exactly quantized in units of $e^2/h$, where $h$ is the Planck constant and $e$ is the elementary charge~\cite{klitzing1980hall}. This phenomenon is called the quantum Hall effect (QHE), always requiring an external magnetic field. In its three-dimensional (3D) version, $\sigma$ is quasi-quantized in units of $e^2/h\cdot k_F^z/\pi$, where $k_F^z$ is the Fermi wave vector of the electrons in the direction of $B$~\cite{tang2019three,galeski2020observation}. 

The same topological invariant is responsible for the intrinsic quantum anomalous Hall effect (QAHE)~\cite{hall1881anomalous,haldane1988model,buhmann2002hgte,PhysRevLett.113.147201,chang2013experimental,checkelsky2014trajectory}. For 2D ferromagnetic insulators, $\sigma$ can be even quantized in units of $e^2/h$ in the absence of an external magnetic field, originating from the intrinsic magnetization of these systems. The 3D equivalent of such quantum anomalous Hall insulators are magnetic Weyl semimetals~\cite{burkov2011weyl}, which are characterized by an anomalous $\sigma$ proportional to $e^2/h\cdot k_D/(2\pi)$, where $k_D$ is the distance between the Weyl nodes~\cite{PhysRevLett.113.187202}. 

Another quantity connected to topological invariants is the thermal Hall conductivity $\kappa$. It is the ratio of a heat current flowing perpendicular in response to a longitudinal temperature gradient under broken TRS~\cite{zhang2000determining}. In comparison to the electronic Hall effect, the thermal Hall effect also contains contributions from neutral excitations and can therefore provide additional, unique information about the topological nature of a state. In 2D electron systems subjected to external magnetic fields, the quantum thermal Hall effect (QTHE) is quantized in units of $(\pi^2k_B^2T)/(3h)$, where $k_B$ is the Boltzmann constant~\cite{kane1997quantized,long2011quantum,sumiyoshi2013quantum,kasahara2018majorana,banerjee2018observation,tikhonov2019quantum}. 

Analogous to $\sigma$ and $\kappa$, the thermoelectric Hall, i.e. Nernst conductivity $\alpha$, is also believed to be quantized~\cite{Nerst_1887,PhysRevB.55.2344,nakamura2005quantum,shirasaki2005impurity,matsuo2009quantum,yang2009thermopower,barlas2012thermopower,PhysRevB.87.075302,PhysRevLett.114.146801}. It is the ratio of a transverse electrical current in response to a longitudinal temperature gradient under broken TRS. In 2D electron systems subjected to external magnetic fields, $\alpha$ is indeed quantized in units of $(ek_B)/h\cdot\ln{2}$~\cite{checkelsky2014trajectory}. This phenomenon is called the quantum Nernst effect (QNE).

For electronic systems, all three, the QHE, the QTHE and the QNE can be understood in terms of a Berry curvature (BC) formalism~\cite{Thouless_1982,Xiao2010,Nagaosa_2010,XiaoDi_2006,PhysRevLett.107.236601,zhang2016berry}. For 2D gapped electron systems with broken TRS, the BC formalism reveals topological invariants called Chern numbers that directly lead to a quantized $\sigma$. However, unlike the QHE, the QTHE and the QNE have so far not been investigated in 2D systems from the BC point of view. Moreover, a generalization to quantum anomalous Hall and 3D systems is also so far elusive. Therefore, it remains an open question whether the thermal Hall and the Nernst effects are as universally quantized for 2D and 3D electron systems with broken TRS as the QHE. 

In this work, we study the Hall effect, the thermal Hall effect and the Nernst effect in the presence of a magnetic field in 2D and 3D electronic systems with both analytical and numerical methods. Moreover, we investigate the anomalous quantum effects in the Haldane model, which we expand to three dimensions. In all cases, we find that $\sigma$ and $\kappa$ are (quasi-)quantized in $B$. However, $\alpha$ is only quantized in 2D systems subjected to an external magnetic field and does not universally express the topological invariant of the investigated systems. Instead, its energy integral is quantized for all cases considered and seems more suitable as thermoelectric quantum than $\alpha$ itself. Our analysis reveals the quantum nature in the thermal Hall and Nernst effect as well as the influence of a weak coupling in the third dimension by utilizing a single universal approach in terms of the BC for all effects.

\section{Models and methods}

For our investigation we employ a tight-binding model on a periodic cubic lattice to ensure that the formalism for bulk systems is applicable. We use the following single-band model to emulate the parabolic dispersion of the free electron gas:
\begin{equation}
 \label{eq:free}
  H^{\text{free}}(\textbf{k})=d(2-(\cos{k_x}+\cos{k_y}))+d_z(1-\cos{k_z}).
\end{equation}
Here, $d$ describes the hopping strength in the $x$-$y$ plane and is linked to the effective mass of the electrons, $m$, via $d=\hbar^2/(2m)$. In analogy to that, $d_z$ gives the hopping strength in the $z$ direction. This separation allows for the description of both the two-dimensional electron gas when setting $d_z=0$ as well as the isotropic three dimensional electron gas with $d_z=d$. Anisotropic cases with a finite $d_z \neq d$ are not investigated within this work. Evaluating this Hamiltonian for low energies and consequently small $k$ allows for the approximation $(1-\cos{k})\approx k^2$ which results in a parabolic effective model:
\begin{equation}
 \label{eq:freeeff}
  H^{\text{free}}_{\text{eff}}(\textbf{k})=d(k_x^2+k_y^2)+d_zk_z^2.
\end{equation}

We implement an external magnetic field $B$ along the $z$ direction utilizing the Peierls substitution~\cite{liang2019gap} with the vector potential $\textbf{A}=(0,Bx,0)$ leading to a modification of the hoppings $d$ as follows
\begin{equation}
\label{eq:peierls}
  d_{\alpha,\beta}\rightarrow d_{\alpha,\beta}e^{i\pi\frac{e}{h}B(x_\alpha+x_\beta)(y_\alpha-y_\beta)}.
\end{equation}
Here, $d_{\alpha,\beta}$ describes the hopping strength from unit cell $\alpha$ located at $x_\alpha$ and $y_\alpha$ to unit cell $\beta$ located at $x_\beta$ and $y_\beta$. The position of the unit cells along $z$ does not appear because the magnetic field $B$ is aligned parallel to $z$. As there is nearest-neighbor interaction only, the hoppings along the $x$ direction remain unchanged due to $y_\alpha-y_\beta=0$. For hoppings along the $y$ direction, the phase factor in equation~\ref{eq:peierls} can be reduced to
\begin{equation}
\label{eq:peierls2}
  d^y_{n}\rightarrow d^y_{n}e^{i\pi\frac{e}{h}B(2na)(a)}.
\end{equation}
Here, $d^y_n$ describes the hopping in the $n$-th unit cell in $x$ direction located at $x=na$, where $a$ is the lattice constant of the underlying grid. Equation~\ref{eq:peierls2} shows, that in general all hoppings $d^y_n$ are different, however, when choosing $B_N=h/(Nea^2)$ with an integer number $N$, the phase factor repeats after $N$ unit cells. Therefore, only magnetic fields obeying this discretization can be evaluated, where each $B_N$ requires a supercell of $N$ cells along the $x$ direction, containing $N$ orbitals. As larger supercells also require more computational effort, this work investigates cases up to $N=400$ for the two-dimensional system and up to $N=100$ for the three-dimensional system. 
\\

The case without an external magnetic field is described by the Haldane model~\cite{haldane1988model}
\begin{align}
\label{eq:haldane}
  H^{\text{Haldane}}(\textbf{k})&=2t_2\cos{\phi}(\sum_{i=1}^3\cos{(\textbf{k}\cdot\textbf{b}_i}))I \nonumber \\&+t_1[\sum_{i=1}^3(\cos{(\textbf{k}\cdot\textbf{a}_i)}\sigma_1+\sin{(\textbf{k}\cdot\textbf{a}_i)}\sigma_2)] \nonumber \\
  &-[M-2t_2\sin\phi(\sum_{i=1}^3\sin{(\textbf{k}\cdot\textbf{b}_i)})]\sigma_3\nonumber\\
  &+d_z\cos{(k_z)}\sigma_3.
\end{align}
Here, the last term is added to introduce a coupling in the third dimension, the two-dimensional case is achieved with $d_z=0$. 

For these TB models we can evaluate the BC as~\cite{Thouless_1982,Xiao2010,Nagaosa_2010}
\begin{equation}
  \Omega_{ij}^n=\sum_{m\ne n} \frac{\langle n|\frac{\partial H}{\partial k_i}|m\rangle \langle m|\frac{\partial H}{\partial k_j}|n\rangle - (i \leftrightarrow j)}{(E_n-E_m)^2}.
\end{equation}
From this, we can utilize the Kubo formula to calculate the intrinsic part of the Hall conductivity as~\cite{Xiao2010,Nagaosa_2010}
\begin{equation}
    \label{eq:ahc}
    \sigma=\frac{e^2}{\hbar}\sum_n \int \frac{d^{n_d}k}{(2\pi)^{n_d}}\Omega^{n\textbf{k}} f_{n\textbf{k}}
\end{equation}
and the intrinsic electronic part of the Thermal Hall conductivity as~\cite{PhysRevLett.107.236601,zhang2016berry}:
\begin{equation}
\label{eq:athc}
 \kappa=\frac{1}{\hbar T}\sum_{n}\int \frac{dk^{n_d}}{(2\pi)^{n_d}}\Omega^{n\textbf{k}} \int_{E_{n\textbf{k}}}^\infty (E-E_F)^2\frac{\partial f}{\partial E}dE.
\end{equation}
The intrinsic Nernst conductivity can be evaluated as~\cite{Xiao2010,XiaoDi_2006}
\begin{align}
\label{eq:anc}
 \alpha&= -\frac{e}{\hbar T} \sum_n \int \frac{d^{n_d}k}{(2\pi)^{n_d}} \Omega^{n\textbf{k}} [(E_{n\textbf{k}}-E_F)f_{n\textbf{k}}\nonumber \\&+k_BT\ln{(1+e^{\frac{E_{n\textbf{k}}-E_F}{-k_BT}})}].
\end{align}
Here, $E_F$ is the Fermi level, $f_{n\textbf{k}}$ is the Fermi distribution, and $n_d$ is the dimensionality of the system. Both Thermal Hall and Nernst conductivity are evaluated for a given temperature $T$.

To calculate these conductivities, an integration in the first Brillouin zone is necessary. Here, for the two-dimensional free electron like model a $k$ mesh of $51\times 51$ is employed. This sparse grid is sufficient because the emerging Landau levels have no dispersion and the Berry curvature is distributed uniformly. However, this is not the case for the three-dimensional expansion of the parabolic band model. Here, along the $k_z$ direction a very dense mesh is needed to correctly capture the dispersive Landau levels, with a $k$ mesh of $51\times 51\times 3001$ showing converged results.

In the Haldane model for two dimensions a $k$ mesh of $301\times 301$ leads to converged results. Here, an expansion into three dimensions is possible with the same spacing, therefore a $301 \times 301\times 301$ mesh was employed.

\section{Analytical results}

For the analytical evaluation of the equations \eqref{eq:ahc}, \eqref{eq:athc}, and \eqref{eq:anc} some pre-requisites are needed. We use the definition of the Chern number in two dimensions as~\cite{berry1984quantal}
\begin{equation}
   \mathcal{C}_n=\frac{1}{2\pi}\int d^2k  \Omega^\textbf{nk},
\end{equation}
where $\Omega^\textbf{nk}$ is the Berry curvature of band $n$ at $\textbf{k}$ in reciprocal space. The Chern number is always an integer in a gapped system.

For a full quantization, the band structure has to have a band gap. When thermal effects are included, this band gap has to be large enough for the respective thermal energy scale $k_BT$. Therefore, we will require for the occupied bands $E_n<<-k_BT$ and for the unoccupied bands $E_n>>k_BT$.
For the analytical derivation, we approximate these expressions with their limiting cases:
\begin{align}
\label{eq:occ}
 \frac{E_n}{k_BT}&<<-1 &\Rightarrow& &\frac{E_n}{k_BT}\rightarrow -\infty \\
\label{eq:unocc}
 \frac{E_n}{k_BT}&>>1 &\Rightarrow& &\frac{E_n}{k_BT}\rightarrow \infty.
\end{align}
These approximations are only valid for $T\rightarrow 0$, therefore we will later use the numerical results to test the validity of the analytical expressions at finite temperatures.

\subsubsection{Useful integrals}

For the evaluation of the conductivities, we will need several integrals that are discussed in the following. For a more detailed discussion of the intermediate steps we refer to the \textit{Supplementary Information}.

The first useful integral is given as
\begin{equation}
\label{eq:aux1}
  \int dx[\frac{xe^x}{e^x+1}+\ln{(e^x+1)}]=x\ln{(e^x+1)}.
\end{equation}
We will also need the limit cases of this function, that are given as
\begin{align}
 \lim_{x\rightarrow\infty} x\ln{(e^x+1)}&\approx x\ln{(e^x)}=x^2 \\
 \lim_{x\rightarrow-\infty} x\ln{(e^x+1)}&\approx x\ln{(1)}=0.
\end{align}
We also need the integral
\begin{equation}
\label{eq:aux2}
 \int dx\ln{(e^x+1)}=-\text{Li}_2(-e^{x}),
\end{equation}
where Li$_2(x)=\sum_{k=1}^\infty \frac{x^k}{k^2}$ is the polylogarithmic function.
We will also need the limit cases of this function, that are given as
\begin{align}
 \lim_{x\rightarrow\infty}\text{Li}_2(-e^x)&=-\frac{x^2}{2}-\frac{\pi^2}{6}+\mathcal{O}(\frac{1}{x^7})\approx-\frac{x^2}{2}-\frac{\pi^2}{6} \\
 \lim_{x\rightarrow-\infty}\text{Li}_2(-e^x)&=0.
\end{align}

We can now calculate the auxiliary integrals $I^I$ as
\begin{equation}
\label{eq:ione}
 I^I=\int dx\frac{x}{e^x+1}=\frac 1 2 x^2 -x\ln{(e^x+1)}-\text{Li}_2(-e^{x})
\end{equation}
and $I^{II}$ as
\begin{align}
\label{eq:itwo}
I^{II}&=\int dx \ln{(1+e^{-x})}\nonumber\\
&=\frac 1 2 x^2-x\ln{(e^x+1)}+x\ln{(e^{-x}+1)}-\text{Li}_2(-e^{x}).
\end{align}

\subsection{2D case}

The following derivations only require the system to have a global band gap, therefore the results are applicable to both the two-dimensional free electron gas model and the Haldane model. The only difference is in the Chern numbers. While in the free electron gas, each Landau level has a Chern number of $\mathcal{C}=1$, in the Haldane model the two bands have opposite Chern numbers. 

\subsubsection{Quantized Hall effect}

For the sake of completeness we start with the quantum (anomalous) Hall effect in two dimensions, which has already been reported in refs.~\cite{klitzing1980hall,haldane1988model}. With the assumption of a global band gap in equations \eqref{eq:occ} and \eqref{eq:unocc}, the Fermi distribution function $f_{n\textbf{k}}$ evaluates to one for occupied and to zero for unoccupied bands, respectively. Evaluating equation \eqref{eq:ahc} then results in the following Hall conductance:
\begin{equation}
 \sigma=\frac{e^2}{\hbar}\sum_n^{occ} \int \frac{d^2k}{(2\pi)^2}\Omega^{n\textbf{k}}=\frac{e^2}{h}\sum_n^{occ}\mathcal{C}_n.
\end{equation}
Because in a gapped system the Chern numbers $\mathcal{C}_n$ are integer, the Hall conductance is quantized in quanta of $\frac{e^2}{h}$.

\subsubsection{Quantized thermal Hall effect}

The thermal Hall conductance results from the following integral (see equation \eqref{eq:athc}):
\begin{equation}
 \kappa=\frac{(k_BT)^2}{\hbar T}\sum_{n}\int \frac{dk^2}{(2\pi)^2}\Omega^{n\textbf{k}} \underbrace{\frac{1}{(k_BT)^2}\int_{E_{n\textbf{k}}}^\infty E^2\frac{\partial f}{\partial E}dE}_{f^\kappa_{n\textbf{k}}}.
\end{equation}
Here, all energy dependent factors are located in the function $f^\kappa_{n\textbf{k}}$ that can be viewed as the weighting of the Berry curvature.

Taking a closer look at $f^\kappa_{n\textbf{k}}$ with the substitution $x=\frac{E}{k_BT}$ leads us to
\begin{align}
 f^\kappa_{n\textbf{k}}=\int_{\frac{E_{n\textbf{k}}}{k_BT}}^\infty x^2\frac{\partial}{\partial x}\frac{1}{e^x+1}dx.
\end{align}
This expression can be further evaluated with $a=\frac{E_{n\textbf{k}}}{k_bT}$ via partial integration to
\begin{align}
  f^\kappa_{n\textbf{k}}=[x^2\frac{1}{e^x+1}]_a^\infty-2\int_a^\infty\frac{x}{e^x+1}dx.
\end{align}
Using the auxiliary integral $I^I$ from equation \eqref{eq:ione}, this results in
\begin{align}
 f^\kappa_{n\textbf{k}}=[x^2\frac{1}{e^x+1}-x^2 +2x\ln{(e^x+1)}+2\text{Li}_2(-e^{x}))]_a^\infty.
 \end{align}
Using the assumption of a large distance of the energy levels from equations \eqref{eq:occ} and \eqref{eq:unocc} for $a$ together with the limit cases discussed in the previous section, one arrives at
\begin{align}
 f^\kappa_{n\textbf{k}}=
        \begin{cases}
            -\frac{\pi^2}{3},& \text{if band $n$ occupied}\\
            0,              & \text{if band $n$ unoccupied}
        \end{cases}
        = -\frac{\pi^2}{3}f_{n\textbf{k}},
\end{align}
where $f_{n\textbf{k}}$ is the Fermi distribution function.

Consequently, in analogy to the quantized Hall conductance a quantization in the electronic thermal Hall conductance emerges for a gapped system as
\begin{align}
 \kappa&=\frac{(k_BT)^2}{\hbar T}\sum_{n}\int \frac{dk^2}{(2\pi)^2}\Omega^{n\textbf{k}} f^\kappa_{n\textbf{k}} \nonumber \\
 &=-\underbrace{\frac{\pi^2k_B^2T}{3h}}_{g_0}\sum_{n}\underbrace{\int \frac{dk^2}{2\pi}\Omega^{n\textbf{k}}}_{\mathcal{C}_n}f_{n\textbf{k}} \nonumber \\
&= -g_0\sum_n^{occ}\mathcal{C}_n
\end{align}
Here, $g_0$ is the thermal conductance quantum. Because in a gapped system the Chern numbers $\mathcal{C}_n$ are integer, the thermal Hall conductance is quantized in quanta of $g_0$.

\subsubsection{Quantized Nernst effect}

Because the Nernst effect obeys the Mott relation and is therefore approximately proportional to the derivative of the Hall effect~\cite{XiaoDi_2006, Xiao2010}, it is dependent on the shape of the Hall conductance in energy. Consequently, the Nernst effect itself can not be quantized. Instead, we utilize the Mott relation to identify a possible quantum value, the integrated Nernst effect $A$. This value is obtained by integrating the Nernst conductance over all occupied states up to the Fermi level $E_F$.
With equation \eqref{eq:anc}, it is possible to express $A$ as
\begin{align}
  A=\int_{-\infty}^{E_F(=0)}dE\alpha(E) &=-\frac{e(k_BT)^2}{\hbar T} \sum_n \int \frac{d^2k}{(2\pi)^2} \Omega^{n\textbf{k}}f^\alpha_{n\textbf{k}} \\
  \text{with   }f^\alpha_{n\textbf{k}}=\frac{1}{(k_BT)^2}&\int_{-\infty}^{0}dE ([(E_{n\textbf{k}}-E)f_{n\textbf{k}}\nonumber\\&+k_BT\ln{(1+e^{\frac{E_{n\textbf{k}}-E}{-k_BT}})}]).
\end{align}
Substituting $x=\frac{E_n\textbf{k}-E}{k_BT}$ and $a=\frac{E_{n\textbf{k}}}{k_bT}$ together with the integrals $I^I$ and $I^{II}$ from equations \eqref{eq:ione} and \eqref{eq:itwo} results in
\begin{align}
 f^\alpha_{n\textbf{k}}&=-\int_{\infty}^{a}dx[\frac{x}{e^x+1}+\ln{(1+e^{-x})}] \nonumber \\
 &=[x^2 -2x\ln{(e^x+1)}+x\ln{(e^{-x}+1)}-2\text{Li}_2(-e^{x})]_a^\infty.
\end{align}
Similarly to the thermal Hall conductance the limit cases can be taken under the assumptions from equations \eqref{eq:occ} and \eqref{eq:unocc} as
\begin{align}
 f^\alpha_{n\textbf{k}}=
        \begin{cases}
            \frac{\pi^2}{3},& \text{if band $n$ occupied}\\
            0,              & \text{if band $n$ unoccupied}
        \end{cases} 
        = \frac{\pi^2}{3}f_{n\textbf{k}},
\end{align}
where $f_{n\textbf{k}}$ is the Fermi distribution function.

As a result, the integrated Nernst conductance can be expressed as:
\begin{align}
 A&=-\frac{e}{\hbar T} \sum_n \int \frac{d^2k}{(2\pi)^2} \Omega^{n\textbf{k}}f^\alpha_{n\textbf{k}} \nonumber \\
 &=-\underbrace{\frac{\pi^2k_B^2T}{3h}}_{g_0}e\sum_{n}\underbrace{\int \frac{dk^2}{2\pi}\Omega^{n\textbf{k}}}_{\mathcal{C}_n}f_{n\textbf{k}} \nonumber \\
&= -g_0e\sum_n^{occ}\mathcal{C}_n
\end{align}
 Here, $g_0$ is the thermal conductance quantum. Because in a gapped system the Chern numbers $\mathcal{C}_n$ are integer, the integrated Nernst conductance is quantized in quanta of $g_0e$.

\subsubsection{Flat bands and the Nernst effect}

While the Nernst conductance itself shows no quantization in general, in the special case of flat bands a quantized value can be found. For example, this is the case for Landau levels in a two-dimensional system, where the bands do not have a dispersion in $\textbf{k}$, i.e. $E_{n\textbf{k}}=E_n$.

In this case, in equation \eqref{eq:anc} the integration in the reciprocal space evaluates to the Chern number and
\begin{align}
  \alpha(E_F)=-\frac{e}{h T} &\sum_n\mathcal{C}_n [(E_{n}-E_F)\frac{1}{e^{\frac{(E_n-E_F)}{k_BT}}+1}\nonumber\\&+k_BT\ln{(1+e^{\frac{(E_{n}-E_F)}{-k_BT}})}].
\end{align}
We can now evaluate the extrema of the Nernst conductivity with respect to $E_F$ via
\begin{align} 
 \frac{\partial \alpha}{\partial E_F}=-\frac{e}{h T} \sum_n\mathcal{C}_n  \frac{\frac{E_n-E_F}{k_BT}(e^\frac{E_n-E_F}{k_BT}+1)}{(e^\frac{E_n-E_F}{k_BT}+1)^2(e^{-\frac{E_n-E_F}{k_BT}}+1)}\overset{!}{=}0.
\end{align}
The derivative can only take a zero value for a Fermi level $E_F^0$ of
\begin{equation}
 \frac{E_n-E_F^0}{k_BT}\overset{!}{=}0\quad\rightarrow\quad E_F^0=E_n.
\end{equation}
From this result we can see, that the extremal value of the Nernst conductance is always located at the energies of the flat bands. There, the corresponding $\alpha^{ext}$ is
\begin{align}
 \alpha^{ext}=-\frac{ek_B}{h}\ln{2}\:\:\mathcal{C}_n.
\end{align}
It is important to note, that this value is, in contrast to the integrated Nernst conductance, not temperature-dependent. This result is consistent with observations from ref.~\cite{PhysRevB.80.081413}. 

\subsection{3D case}

\subsubsection{Free electron gas}

For the three-dimensional free electron gas, a magnetic field still leads to the emergence of Landau levels. However, while they are flat in the plane perpendicular to the field, the dispersion in the direction along the field is not changed. For a magnetic field $B$ along $z$, the dispersion is given as
\begin{equation}
\label{eq:3ddisp}
 E_{n\textbf{k}}=n\frac{e\hbar}{m_e}B+dk_z^2.
\end{equation}
The fact, that the band dispersion in the $k_x-k_y$ plane remains unchanged from the 2D case, can be utilized to simplify the calculation of the three-dimensional integrals for the conductivities: The integral can be separated in two parts and the two-dimensional result from the previous section can be utilized. For the Hall conductivity, this results in the following expression.\begin{align}
  \sigma &=\frac{e^2}{\hbar}\sum_n \int\frac{dk_z}{2\pi}\iint\frac{dk_xdk_y}{(2\pi)^2}f_{n\textbf{k}} \nonumber \\
  &=\frac{e^2}{h}\sum_n\mathcal{C}_n \int\frac{dk_z}{2\pi}\begin{cases}
        1, &E_{nk_z}<E_F \\
        0, &E_{nk_z}>E_F 
  \end{cases}
        \nonumber \\
  &=\frac{e^2}{h}\sum_n\mathcal{C}_n \int\frac{dk_z}{2\pi}\begin{cases}
        1, &|k_z|<\sqrt{\frac{E_F}{d}-n\frac{e\hbar}{m_e}\frac{B}{d}} \\
        0, &|k_z|>\sqrt{\frac{E_F}{d}-n\frac{e\hbar}{m_e}\frac{B}{d}}
  \end{cases}
        \nonumber \\
  &=\frac{e^2}{h}\frac{1}{\pi}\sum_n\mathcal{C}_n \sqrt{\frac{E_F}{d}-n\frac{e\hbar}{m_e}\frac{B}{d}}.
\end{align}

While no robust quantization can be found from this expression, it is still interesting to investigate the values corresponding to the Landau level onsets.

To simplify our results, we define a $k_0$ in the way to describe the Fermi wave vector in $k_z$ direction, when the second Landau level crosses the Fermi energy:
\begin{equation}
 \frac{e\hbar}{m_e}B=dk_0^2 \rightarrow k_0=\sqrt{\frac{e\hbar}{m_e}\frac{B}{d}}.
\end{equation}
 Evaluating the value at the onset of the $m-th$ Landau level together with $\mathcal{C}_n=1$ results in
\begin{align}
\label{eq:3dhall}
 \sigma=\frac{e^2}{h}\frac{k_0}{\pi}\sum_{k=1}^{m} \sqrt{k}.
\end{align}
Here, the three-dimensional value is derived from the two-dimensional conductance quantum scaled by a factor, which only involves an intrinsic parameter of the electronic system, $k_0$. Therefore, the three-dimensional Hall conductivity can be viewed as quasi-quantized.

In analogy to the calculation above, we find similar results for the thermal Hall and integrated Nernst conductivities
\begin{align}
 \kappa&=-g_0\frac{k_0}{\pi}\sum_{k=1}^{m} \sqrt{k}\\
 A&=-g_0e\frac{k_0}{\pi}\sum_{k=1}^{m} \sqrt{k}.
\end{align}
Like for the Hall conductivity, the three-dimensional values are derived from the two-dimensional conductance quanta scaled by the same factor depending only on $k_0$.

It is important to note, that the scaling wavelength $k_0$ is not equivalent to the Fermi wavelength $k_F$. Instead, they a connected via
\begin{align}
 E_F&=d{k_F^z}^2=\frac 3 2\frac{e\hbar}{m_e}B=\frac 1 2\frac{e\hbar}{m_e}B+dk_0^2\nonumber\\
 &\rightarrow k_0=\sqrt{\frac{e\hbar}{m_eB}\frac 1 d}=\sqrt{\frac 2 3 \frac{E_F}{d}}=\sqrt{\frac 2 3}k_F^z.
\end{align}

We also evaluate the extremal Nernst conductivity analogously to the two-dimensional case with the dispersion from equation \eqref{eq:3ddisp}. Here, the integral over $k_z$ is not analytically solvable, but can be treated numerically. This results in the following expression
\begin{equation}
 \alpha_{ext}\approx-\frac{ek_B}{h}\ln{2}\sqrt{\frac{k_BT}{d}}*0.548.
 \end{equation}
While this value is not as clearly connected to the two-dimensional value as the other quantities, it nevertheless only depends on the temperature $T$ and the material parameter $d$ which is linked to the effective mass. This value gives only the height of the first peak in the Nernst conductivity, as the following peaks are always superpositions of several Landau levels.

\subsubsection{Weyl semimetal}

The three-dimensional expansion of the Haldane model leads to an ideal Weyl semimetal with Weyl points as linear point crossings of bands and no other bands at the Fermi level. In this case we can divide the three-dimensional Brillouin zone of the system in two-dimensional slices along $k_z$. Each of these slices either contains a Weyl point or has a band gap. For all slices that are gapped, we can use our results from the previous section to find quantized conductivities according to the Chern number $\mathcal{C}_n^{k_z}$ of the slice (which can also be zero). Because the Chern number only changes when the band gap closes, all slices have the same Chern number until a Weyl point is crossed. In this case the Chern number is changed by one. Here, the sign is dependent on the chirality of the Weyl point.\\

At a Weyl point with chirality $c$ the Chern number of the 2D Haldane models changes by $-c$. Assuming that there are $N$ Weyl points with chiralities $c_i$ located at $k_z^i$, the Chern number between Weyl point $i$ and $i+1$ is therefore $\mathcal{C}_n=-\sum_{j=1}^{i}c_j$. Additionally, because Weyl points always exist in pairs, we know $\sum_{j=1}^{N}c_j=0$. In the following we assume for simplicity to have only one occupied band. In this case, we get for the Hall conductivity
\begin{align}
  \sigma&=\frac{e^2}{h}\int\frac{dk_z}{2\pi}\mathcal{C}^{k_z} \nonumber \\
  &=\frac{e^2}{h}\frac{1}{2\pi}\sum_{i=1}^{N-1}\left[(k_z^i-k_z^{i+1})\sum_{j=1}^{i}c_j\right].
\end{align}
By transforming the sums this can be further evaluated as
\begin{align}
  \sigma=\frac{e^2}{h}\frac{1}{2\pi}\sum_{i=1}^{N}k_z^ic_i.
\end{align}
We can define a generalized Weyl point distance $k_D$ as
\begin{equation}
 k_D=\sum_{i=1}^{N}k_z^ic_i,
\end{equation}
which reduces to the distance of the Weyl points for a single pair of points. With this generalized distance we get
\begin{equation}
 \sigma=\frac{e^2}{h}\frac{k_D}{2\pi}.
\end{equation}
This result is analogous to the first step ($m=1$) of the three-dimensional free electron gas from equation \eqref{eq:3dhall}. Similarly, we find a value which is not universally quantized, but consists of the two-dimensional conductance quantum scaled by a characteristic wavelength. Therefore this result can be viewed as quasi-quantized. For the simplest case of two Weyl points, our result is consistent with the one reported in ref.~\cite{PhysRevLett.113.187202}.

With the same idea of treating the slices of the Brillouin zone separately we also get quasi-quantized thermal Hall and integrated Nernst conductivities as
\begin{align}
 \kappa&=-g_0 \frac{k_D}{2\pi} \\
 A&=-g_0e \frac{k_D}{2\pi}.
\end{align}
Also here, both values are given by the respective two-dimensional quanta scaled by the same characteristic wavelength.

\subsection{Summary of the calculated (quasi-)quanta}

\begin{table}
 \begin{tabular}{|c|c|c|c|c|}
\hline
        &   $\sigma$    &    $\kappa$   &   $\alpha^{ext}$      & $A$ \\ \hline
2D free &  $\frac{e^2}{h}$  &  $-g_0$    &  $-\frac{ek_B}{h}\ln{2}$  &  $-g_oe$ \\ \hline
3D free & $\frac{e^2}{h}\frac{k_0}{\pi}$ & $-g_0\frac{k_0}{\pi} \sqrt{k}$ &  $-\frac{ek_B}{h}\ln{2}\sqrt{\frac{k_B T}{d}}0.548$ & $-g_0e\frac{k_0}{\pi}$ \\ \hline
2D Haldane &  $\frac{e^2}{h}$  &  $-g_0$    &  $n.a.$  &  $-g_0e$ \\ \hline
3D Haldane & $\frac{e^2}{h}\frac{k_D}{2\pi}$ & $-g_0 \frac{k_D}{2\pi}$ & $n.a.$ & $-g_0e \frac{k_D}{2\pi}$  \\ \hline
 \end{tabular}
\label{tab:quanta}
\caption{Overview of the quanta. Shown are the calculated quanta for the Hall conductivity $\sigma$, the thermal Hall conductivity $\kappa$, the extremal Nernst conductivity $\alpha^{ext}$, and the integrated Nernst conductivity $A=\int_{-\infty}^{E_F}dE\alpha(E)$. $k_0$ is the characteristic wave vector in the direction of the magnetic field and $k_D$ is the generalized Weyl point distance. $g_0=(\pi^2k_B^2T)/(3h)$ is the thermal conductance quantum. It is important to note, that $\alpha^{ext}$ is only quantized in the free electron gas.}
\end{table}

The calculated values for Hall, thermal Hall, integrated Nernst, and extremal Nernst conductances and conductivities exhibit a lot of similarities for both the normal case of a free electron gas in an external magnetic field and the anomalous case without external fields. An overview of the calculated two-dimensional quanta and their related three-dimensional quasi-quanta is given in Tab. \ref{tab:quanta}. We find, that the more general quantum expression for the Nernst conductivity is obtained from the integrated value, while the extremal part is only important for flat band systems. Additionally, the analogy between the three-dimensional Hall and anomalous Hall systems becomes obvious: In both cases the obtained value consists of the two-dimensional conductance quantum scaled by a characteristic wavelength, which results from the electronic structure.

\section{Numerical results}

In the following, we discuss the general behaviour of the models from equations \eqref{eq:free} and \eqref{eq:haldane} and evaluate them numerically for large parameter sets to check the analytical predictions and estimate the validity of the approximations from equations \eqref{eq:occ} and \eqref{eq:unocc}.

\subsection{Free electron gas model}

\subsubsection{Two-dimensional case}

\begin{figure*}[htb]
\centering
\includegraphics[width=\textwidth]{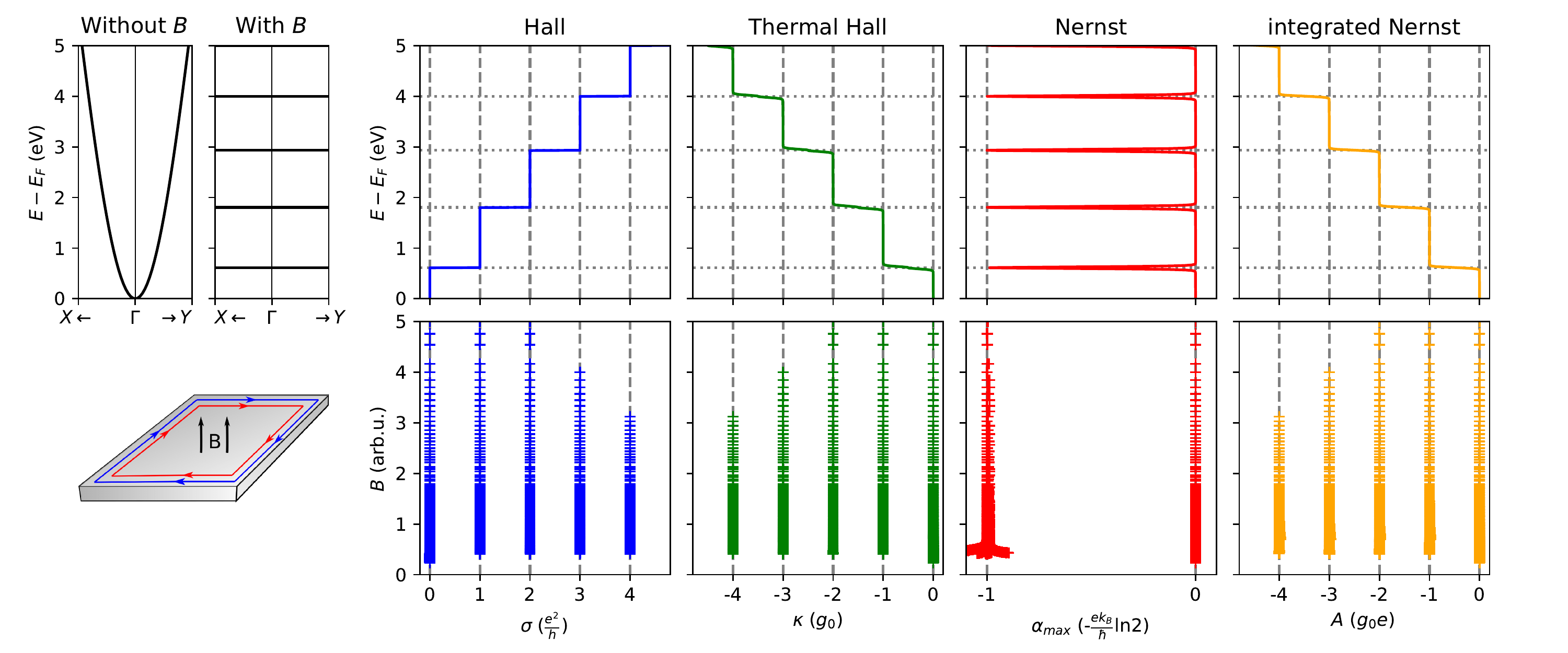}
   \caption{The 2D free electron gas. Upper left:  Band structure without and with magnetic field $B$. Lower left: Schematics of the 2D Quantum Hall effect. Upper right: Hall, thermal Hall, Nernst, and integrated Nernst conductance with Fermi level $E_F$. The horizontal dotted lines give the energy position of the Landau levels. Exemplarily plotted shown for $d=3$ eV and $n=30$. Lower right: Numerical results showing the quantization in the two-dimensional free electron gas. All values are plotted against the magnetic field. Hall plateaus are quantized in electrical conductance quanta $\frac{e^2}{h}$. Thermal Hall plateaus are quantized in thermal conductance quanta $g_0=\frac{\pi^2k_B^2T}{3h}$. Extremal Nernst effect shows a single peak height of $\frac{ek_B}{h}\ln{2}$. The deviation at small $B$ is due to the small energy difference between the Landau levels. Integrated Nernst plateaus are quantized in quanta of $g_0e$.}
\label{fig:free2D}
\end{figure*}

%

For the free electron gas, in Fig.~\ref{fig:free2D} the effect of the magnetic field on the band structure is shown. Here, the parabolic bands in the low energy range become completely flat Landau levels. Evaluating now $\sigma$, $\kappa$, and $\alpha$ for this case, we find a step-like behavior in both Hall, thermal Hall, and integrated Nernst conductance $A$ (see Fig.~\ref{fig:free2D}). Each plateau between the steps is exactly a multiple of the quantization value found in the analytical analysis and the steps occur always when a Landau level crosses the Fermi energy. $\alpha$ is always 0 except at the Landau level crossing where it reaches the quantized value found from the analytical evaluation. 


We have calculated the two-dimensional free electron gas model with $d=2,3,4,5,10$ eV and $n$ ranging from $1$ to $400$. We extracted for each parameter set the height of the Hall, thermal Hall, and integrated Nernst conductance plateaus and the height of the peaks in the Nernst conductance. The values are shown in Fig.~\ref{fig:free2D}. All values are integer multiples of the calculated quanta, in the extremal Nernst effect the peak height is always one quantum. The deviation at small fields is due to the fact, that the assumption of widely separated energy levels is no longer valid and therefore the numerical result is a superposition of the peaks from several Landau levels.

\subsubsection{Three-dimensional case}

\begin{figure*}[htb]
\centering
\includegraphics[width=\textwidth]{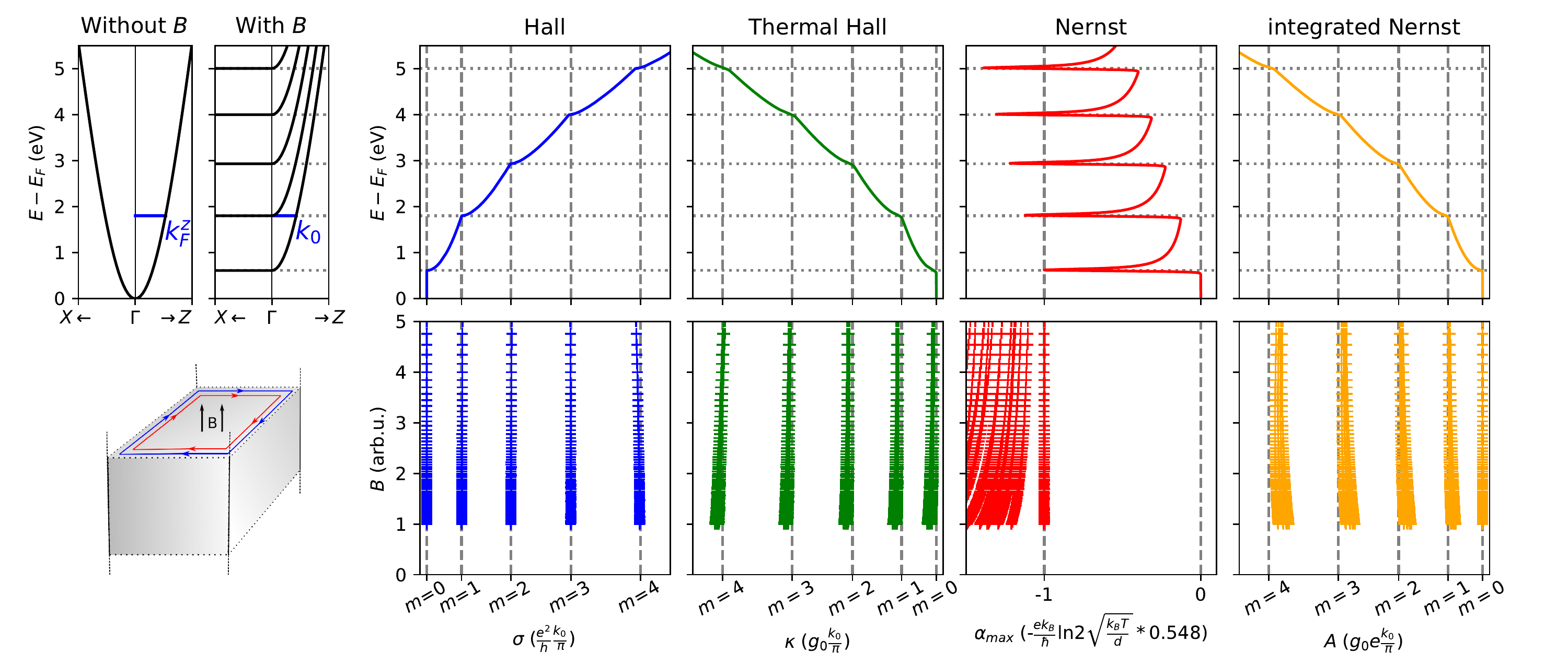}
   \caption{The 3D free electron gas. Upper left:  Band structure without and with magnetic field $B$. Without $B$ the bands are parabolic for low energies with Fermi wave vector $k_F^z$, with $B$ there are flat Landau levels for constant $k_z$ and a parabolic dispersion along $k_z$ with the characteristic wave vector $k_0$. Lower left: Schematics of the 3D Quantum Hall effect. Upper right: Hall, thermal Hall, Nernst, and integrated Nernst conductivity with Fermi level $E_F$. The kinks occur when crossing a Landau level (dotted horizontal lines). $\sigma$, $\kappa$, and $A$ are quantized with $k_0/\pi$ as a scaling factor. The first peak of $\alpha$ is also quantized. Exemplarily plotted shown for $d=d_z=3$ eV and $n=30$. Lower right: Numerical results showing the quantization in the three-dimensional free electron gas. All values are plotted against the magnetic field and extracted at the kinks. Hall plateaus are quasi-quantized in scaled electrical conductance quanta $\frac{e^2}{h}\frac{k_0}{\pi}\sum_{i_1}^m\sqrt{i}$. Thermal Hall plateaus are quasi-quantized in scaled thermal conductance quanta $g_0\frac{k_0}{\pi}\sum_{i_1}^m\sqrt{i}$. Extremal Nernst effect shows a quasi-quantized first peak height of $\frac{ek_B}{h}\ln{2}\sqrt{\frac{k_BT}{d}}*0.548060323$. The deviation at small $B$ is due to the small energy difference between the Landau levels. Integrated Nernst plateaus are quasi-quantized in scaled quanta of $g_0e\frac{k_0}{\pi}\sum_{i_1}^m\sqrt{i}$. For the higher Landau levels (larger $m$) and larger magnetic fields the results deviate because the energy of these Landau levels is too large to be still in the parabolic regime of the cosine function.}
\label{fig:free3D}
\end{figure*}


As shown in Fig.~\ref{fig:free3D}, in the three-dimensional case the magnetic field creates flat Landau levels in the $k_x-k_y$ plane that have a quadratic dispersion along the $k_z$ direction. Both Hall, thermal Hall, and integrated Nernst conductivity show a similar behaviour. They start at the energy of the first Landau level at $k_z=0$ and increase with $\sqrt{E}$ (see Fig. \ref{fig:free3D}). At the energy of the second Landau level, an additional square-root function starts and is added to the conductivities. At the onsets of the Landau levels there are characteristic kinks, that have a value of the 2D quanta scaled by the characteristic wavevector $\frac{k_0}{\pi}$ and a factor of $\sum_{k=1}^{m} \sqrt{k}$ to take into account lower lying Landau levels. The Nernst conductivity shows peaks just like in the 2D case but now these peaks have slowly decaying tails to higher energies because all Landau levels disperse infinitely in $k_z$. Therefore, only the first peak has a predictable value.


We have calculated the three-dimensional electron gas model with $d=3,4,5$ eV and $n$ going from $1$ to $100$. For the evaluation we extracted the values of Hall, thermal Hall, and integrated Nernst conductivity at the kinks, which correspond to the onset energy of the Landau levels at $k_z=0$. As it can be seen in Fig.~\ref{fig:free3D}, the numerical results correspond very good to the analytical predictions. The deviation for larger fields and higher Landau levels is because the parabolic approximation fails at these energies. For the integrated Nernst conductivity there is an additional uncertainty, because it is not possible to separate the influence of the different Landau levels. Due to the temperature broadening, the Nernst conductivity of a certain Landau level is already present slightly below it and consequently, the integral includes a small error. The extremal Nernst conductivity of the first peak is also exactly quantized, for higher peaks this is not the case.

\subsection{Haldane model}

\subsubsection{Two-dimensional case}

\begin{figure*}[htb]
\centering
\includegraphics[width=\textwidth]{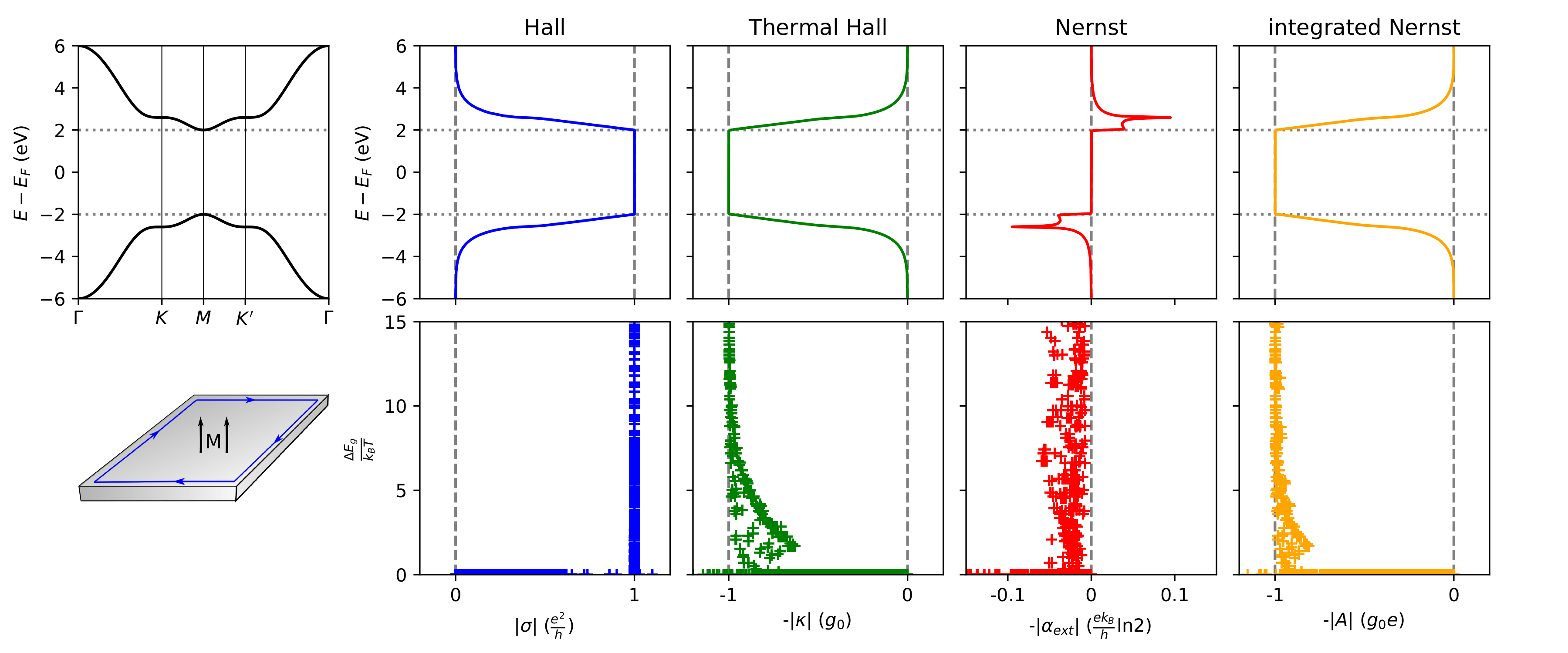}
   \caption{The 2D Haldane model. Upper left:  Band structure. Lower left: Schematics of the 2D Quantum anomalous Hall effect. Upper right: Hall, thermal Hall, Nernst, and integrated Nernst conductance with Fermi level $E_F$. The dotted horizontal lines mark the global band gap. $\sigma$, $\kappa$, and $A$ are quantized in the band gap. $\alpha$ has no quantization. Exemplarily shown for $M=0$ eV, $\phi=0.5\pi$, $t_1=2$ eV, and $t_2=0.5$ eV. Lower right: Numerical results showing the quantization in the two-dimensional Haldane model. All values are plotted against the band gap over $k_BT$. Anomalous Hall conductance is quantized with the electrical conductance quantum $\frac{e^2}{h}$. (b) Anomalous thermal Hall conductance is quantized with the thermal conductance quantum $g_0=\frac{\pi^2k_B^2T}{3h}$. (c) Extremal Anomalous Nernst effect is not quantized. (d) Integrated Anomalous Nernst conductance is quantized with $g_0e$. For small $\frac{\Delta E_G}{k_BT}$ the quantization value is not reached due to thermal excitations.}
\label{fig:haldane2D}
\end{figure*}
%

The band structure shown in Fig.~\ref{fig:haldane2D} shows the inverted band gap of the two-dimensional Haldane model. It can be seen, that both conduction and valence band are indeed not flat, which is in contrast to the Landau levels discussed for the two-dimensional free electron gas. The conductances shown in Fig.~\ref{fig:haldane2D} reach the predicted quantized value in the band gap. It can also be seen, that as a consequence of the dispersive bands $\alpha$ is streched in energy. The exact shape is hereby determined by the dispersion. Consequently, the $\alpha_{ext}$ itself is material dependent, but $A$ remains a quantized value. 
%

We have calculated the two-dimensional Haldane model for parameter sets with $m$ ranging from -6 to 6 eV in 51 steps, $\phi$ ranging from 0 to $2\pi$ in 51 steps, $t_1=-2,-1,1,2$ eV and $t_2=0.5$ eV. We extracted the Hall, thermal Hall and integrated Nernst conductances in the middle of the band gap and the extremal Nernst conductance from the lower half of the band structure. In Fig.~\ref{fig:haldane2D} the extracted values are shown in dependence of the band gap size over the thermal energy scale $k_BT$. For larger band gaps the conductances reach the analytically predicted quanta. For small band gaps thermal excitations are larger than the energy gap, leading to not fully quantized values. With the help of these results, we can review the approximation from the analytical evaluation in equations~\ref{eq:occ} and \ref{eq:unocc}. From this we find, that for fully quantized values the band gap has to satisfy the relation $\Delta E_G>10 k_BT$.

\subsubsection{Three-dimensional case}
\begin{figure*}[htb]
\centering
\includegraphics[width=\textwidth]{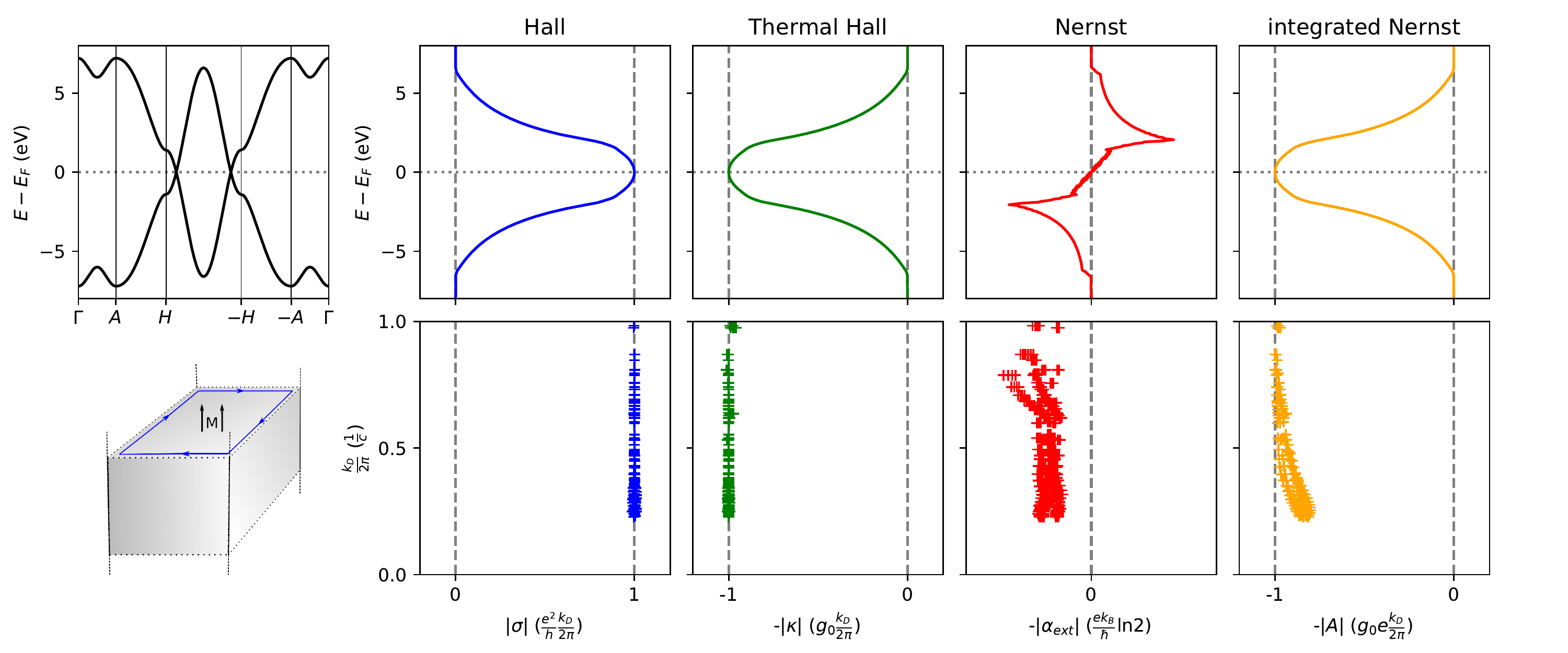}
   \caption{The 3D Haldane model. Upper left:  Band structure. Lower left: Schematics of the 3D Quantum anomalous Hall effect. Upper right: Hall, thermal Hall, Nernst, and integrated Nernst conductivity with Fermi level $E_F$. $\sigma$, $\kappa$, and $A$ reach the 2D quantum scaled by the generalized Weyl point distance $k_D/(2\pi)$ at the Weyl point energy (dotted horizontal line). $\alpha$ has no quantization. Exemplarily shown for $M=0$ eV, $\phi=0.5\pi$, $t_1=2$ eV, $t_2=0.5$ eV, and $d=2$ eV. Lower right: Numerical results showing the quantization in the three-dimensional Haldane model. All values are plotted against the effective weyl point distance $\frac{k_D}{2\pi}$. Anomalous Hall conductivity is quasi-quantized. Anomalous thermal Hall conductivity is quasi-quantized. Extremal Anomalous Nernst effect does not scale. Integrated Anomalous Nernst conductivity is quasi-quantized. The small deviation comes from the fact, that very close to the Weyl points the assumption of a large enough band gap is no longer valid.}
\label{fig:haldane3D}
\end{figure*}
%

To investigate the Haldane model in three dimensions we add a coupling of the Haldane layers in the third dimension. For small coupling parameters, the system stays insulating, while increasing the interactions leads to the emergence of a pair of Weyl points~\cite{burkov2011weyl}. Further increasing it also increases the number of Weyl points in the system. In Fig.~\ref{fig:haldane3D} the band structure of a system with Weyl points is shown. Looking at the conductivities in Fig.~\ref{fig:haldane3D} there is no longer a quantized plateau because the band gap is closed at the Weyl points. Nevertheless at the Weyl point energy the values are exactly the 2D quanta scaled by $k_D/(2\pi)$ for $\sigma$, $\kappa$ and $A$. $\alpha$ itself shows no quantization.

%

We have calculated the three-dimensional Haldane model with the following parameter sets: $m=-1,0,1$ eV, $\phi=0.3\pi,0.5\pi,0.7\pi$, $t_1=2$ eV, $t_2=0.5$ eV, and $d$ ranging from 0 to 5 in 101 steps. We extracted the Hall, thermal Hall, and integrated Nernst conductivity at the energy of the Weyl points. In Fig.~\ref{fig:haldane3D} they are plotted against the generalized Weyl point distance $\frac{k_D}{2\pi}$ and a linear relation is visible for all three values. The three-dimensional quasi-quantization is therefore the 2D quantum scaled by this factor. In the integrated Nernst conductivity there are deviations from the linear relation for a small Weyl point distance. This is because if the Weyl points are close together, the band gap is not very large between them. Therefore, the assumption of $\Delta E_G>>k_BT$ is not valid in a part of the Brillouin zone, leading to a deviation from the predicted value. In the extremal Nernst conductivity there is no quantization.

\section{A Remark on the Quantum Spin Hall effect}

In the Spin Hall effect there is no transverse charge current but a transverse spin current~\cite{PhysRevLett.83.1834}. For the Quantum Spin Hall effect (QSHE) in two dimensions this spin current becomes quantized, which has been realized experimentally by M. K{\"o}nig \textit{et al.}~\cite{konig2007quantum}. A possible model for the QSHE consists of two copies of the Haldane model which are the conjugate of each other~\cite{kane2005quantum}. In this kind of model everything shown for the Haldane model still holds, leading to similar (quasi-)quantizations in the QSHE and the Quantum Spin Nernst effect in both two and three dimensions.

\section{A Remark on the quasi-quantization of the three-dimensional conductivities}

For the two-dimensional systems, our calculations reveal the existance of conductance quanta for the investigated effects that agree with the already reported results about the separate effects. However, the results for the three-dimensional conductivity values do not show a rigorous quantization. Nevertheless, as these values are given by the respective two-dimensional conductance quanta scaled by a characteristic wavelength, they can be viewed as quasi-quantized. The values only depend on an intrinsic property of the bulk band structure. Therefore, local perturbations or disorder effects, that do not influence the band structure globally, do not change the quasi-quantized conductivities. In that sense, they can be viewed as topologically stabilized and robust against local perturbations.

\section{Summary}

In summary, we investigated both numerically and analytically the quantization of Hall, thermal Hall, and Nernst effect in a free electron gas model and the Haldane model for two and three dimensions. In agreement with previous reports, we find that the Hall and thermal Hall effects in two dimensions are quantized both for an external magnetic field and an internal magnetization. To achieve a universal quantized value in a thermoelectric quantity it is neccessary to integrate the Nernst conductance in energy. Additionally, when the bands are completely flat, e.~g. in Landau levels, the extremal Nernst conductance becomes quantized. While in two dimensions all these effects are quantized with natural constants, our calculations in three dimensions reveal, that these 2D quanta are scaled by a characteristic length, namely the Fermi wave vector of the first Landau level for the free electron gas and the generalized Weyl point distance in the Haldane model. Therefore, we conclude the three-dimensional conductivities to be quasi-quantized. 


Our results reveal a universal (quasi-)quantization of thermal and thermoelectric transport in two and three dimensions from a single approach using the Berry curvature for all three effects and show the common origin of both normal and anomalous transport effects.

This work was financially supported by ERC Advanced Grant No. 742068 'TOPMAT' and EC project FET-OPEN SCHINES No. 829044.

\end{document}